\documentstyle[aps,tighten,epsf,amsmath]{revtex}

\begin{document}

\draft

\title{A classical Over Barrier Model to compute charge exchange 
between ions and one--optical--electron atoms }

\author{Fabio Sattin \thanks{E-mail: sattin@igi.pd.cnr.it}}

\address{Consorzio RFX, Corso Stati Uniti 4, 35127 Padova, ITALY}

\maketitle

\abstract{
In this paper we study theoretically the process of electron capture between 
one--optical--electron atoms (e.g. hydrogenlike or alkali atoms) and ions at low-to-medium 
impact velocities ($v/v_e \leq 1$) working on a modification of an already developed classical 
Over Barrier Model (OBM) [V. Ostrovsky, J. Phys. B: At. Mol. Opt. Phys. {\bf 28}
3901 (1995)], which allows to give a semianalytical formula for the cross sections.
The model is discussed and then applied to a number of test cases including experimental data 
as well as data coming from other sophisticated numerical simulations.
It is found that the accuracy of the model, with the suggested corrections and applied to quite 
different situations, is rather high. }
 

\section{Introduction}
The electron capture process in collisions of slow, highly charged 
ions with neutral atoms and molecules is of great importance not only 
in basic atomic physics but also in applied fields such as fusion 
plasmas and astrophysics. The process under study can be 
written as:
\begin{equation}
\label{eq:collisioni1}
A^{+q} + B  \to A^{(q-j)+} + B^{j+} \quad .
\end{equation}
Theoretical models are regularly developed and/or improved to solve 
(\ref{eq:collisioni1}) from first principles for a variety of choices of target $A$ and
the projectile $B$, and their predictions are compared with the results of ever
more refined experiments.\\
In principle, one could compute all the quantities of interest by writing
the time-dependent Schr\"odinger equation for the system (\ref{eq:collisioni1})
and programming a computer to solve it. 
This task can be performed on present--days supercomputers for moderately 
complicated systems. 
Notwithstanding this, simple approximate models are still valuable:
(i) they allow to get analytical estimates which are easy to adapt to 
particular cases; 
(ii) allow to get physical insight on the features of the problem by looking at
the analytical formulas; 
(iii) finally, they can be the only tools available when the complexity 
of the problem overcomes the capabilities of the computers. 
For this reason new models are being still developed \cite{ostrovsky,nove,jpb}.

The present author has presented in a recent paper \cite{jpb} a study 
attempting to develop a more accurate OBM by adding some quantal 
features. The model so developed was therefore called a 
semi--classical OBM.  Its results showed somewhat an improvement with 
respect to other OBMs, but not a dramatic one.\\ 
In this paper we aim to present an OBM for dealing with one of the simplest
processes (\ref{eq:collisioni1}): that between an ion and a target provided
with a single active electron. Unlike the former one \cite{jpb}, this 
model is entirely developed within the framework of a classical 
model, previously studied in \cite{ostrovsky} (see also 
\cite{ryufuku}), but with some important amendments and improvements 
which, as we shall see, allow a quite good accordance with experiments. 

The paper is organized as follows:
a first version of the model is presented and discussed 
in section II. In section III we will test our model against a first test case.   
From the comparison a further improvement
to the model is proposed (section IV) and tested against the same case, as well as 
other data in section V. It will be shown that predictions with this correction
are in much better agreement.

\section{The model: first picture}
We consider the standard scattering experiment and label {\bf T}, 
{\bf P}, and {\bf e} respectively the target ion, the projectile and the
electron. The system {\bf T} + {\bf e} is the initial neutral atom. Let {\bf r} 
be the electron vector relative to {\bf T} and {\bf R} the internuclear 
vector between {\bf T} and {\bf P}. In the spirit of classical OBM models, all particles
are considered as classical objects. \\ 
Let us consider the plane ${\cal P}$ containing all the three particles and use 
cylindrical polar coordinates $(\rho, z, \phi)$ to 
describe the position of the electron within this plane. We can arbitrarily choose
to set the angle $\phi = 0$, 
and assign the $z$ axis to the direction along the internuclear axis. \\ 
The total energy of the electron is (atomic units will be used unless 
otherwise stated):
\begin{equation}
\label{eq:uno}
E = {p^{2} \over 2 } + U = {p^{2} \over 2 }
-{ Z_{t} \over \sqrt{\rho^{2}+z^{2} } }- {Z_{p} \over \sqrt{\rho^{2}+(R-z)^{2}}} 
\quad .
\end{equation}
$Z_{p}$ and $Z_{t}$ are the effective charge of the projectile and  
of the target seen by the electron, respectively. Notice that we are 
considering hydrogenlike approximations for both the target and the 
projectile. We assigne an effective charge $Z_t = 1$ to the target and
an effective quantum number $n$ to label the binding energy of the 
electron:$ E_{n} = Z_{t}^{2}/ 2 n^{2} = 1/ 2 n^{2}$.

As long as the electron is bound to {\bf T}, we can also approximate $E$ as 
\begin{equation}
\label{eq:due}
E(R) = - E_{n} - {Z_{p} \over R} \quad .
\end{equation}
This expression is used throughout all calculations in 
(I); however, we notice that it is asimptotically correct as long as as $ R \to \infty$.
In the limit of small $R$, instead, $E(R)$ must converge to a finite 
limit: 
\begin{equation}
\label{eq:duefinito}
E(R) \to (Z_{p} + 1)^{2} E_{n} 
\end{equation}
(united atom limit).  For the 
moment we will assume that $R$ is sufficiently large so that eq . 
(\ref{eq:due}) holds, but later we will consider the limit 
(\ref{eq:duefinito}), too.
 
\begin{figure}
\epsfxsize=12cm
\epsfbox{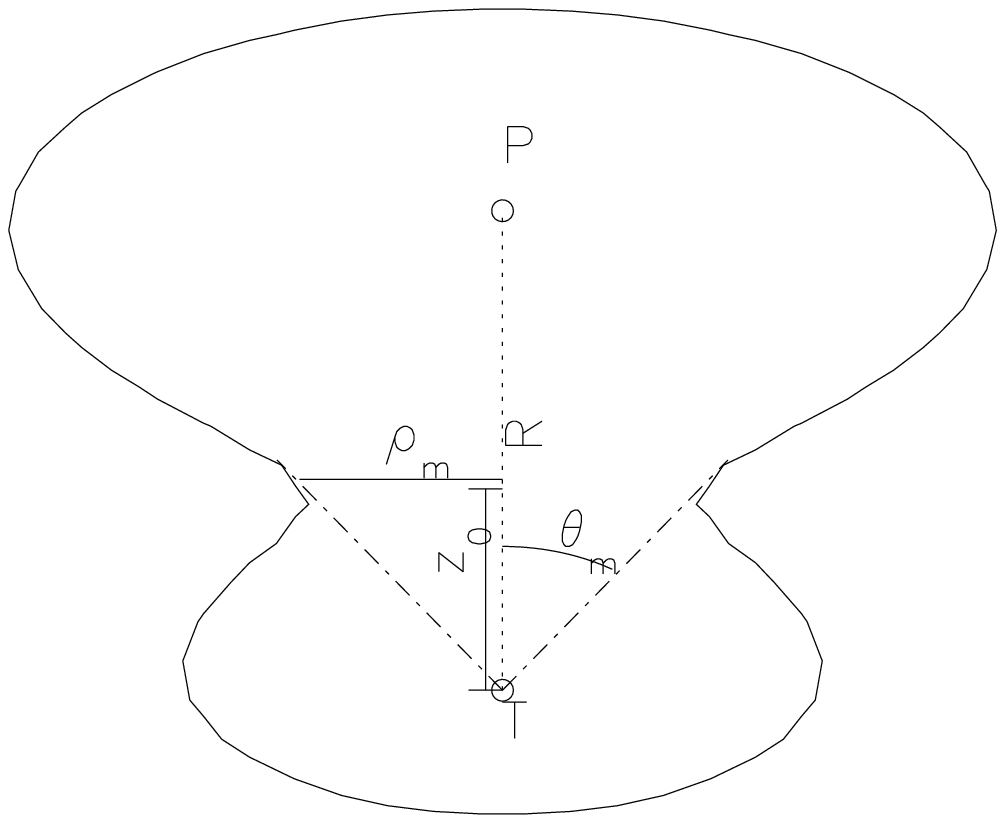}
\caption{The enveloping curve shows 
a section of the equipotential surface $U = E$, i.e. it is the border 
of the region classically accessible to the electron. $R$ is the 
internuclear distance. The parameter 
$\rho_{m}$ is the radius of the opening which joins the potential 
wells, $\theta_{m}$ the opening angle from {\bf T};
$z_{0}$ is the position of the potential's saddle point.}
\label{fig:opening}
\end{figure}

On the plane ${\cal P}$ we can draw a section of the equipotential surface 
\begin{equation}
\label{eq:equip}
 U(z,\rho,R) = - E_n  - {Z_p \over R} \quad . 
\end{equation}
This represents the limit of the region classically allowed to the electron.
When $R \to \infty$ this region is divided into two 
disconnected circles centered around each of the two nuclei. Initial
conditions determine which of the two regions actually the electron lives in.
As $R$ diminishes there can be eventually an instant where the two regions
become connected. In fig. \ref{fig:opening} we give an example for this. \\
In the spirit of OBMs it is the opening of
the equipotential curve between {\bf P} and {\bf T}
which leads to a leakage of electrons from one nucleus to another, and
therefore to charge exchange.
We make here the no-return hypothesis: once crossed the barrier, the electron does 
not return to the target. It is well justified if $Z_p >> 1$. As we shall see just below,
this hypothesis has important consequences. 
 
It is easy to solve 
eq. (\ref{eq:equip}) for $R$ by imposing a vanishing width of the opening 
($\rho_{m} = 0$); furthermore, by imposing also that there be an unique solution 
for $z$ in the range $ 0 < z < R$:
\begin{equation}
\label{eq:rm}
R_{m} = { (1 + \sqrt{Z_{p}})^{2} - Z_{p} \over E_{n} } \quad .
\end{equation} 
In the region of the opening the potential $U$ has a saddle structure:
along the internuclear axis it has a maximum at
\begin{equation}
\label{eq:saddle}
z = z_{0} = R { 1 \over \sqrt{Z_{p}} + 1 } 
\end{equation} 
while this is a minimum along the orthogonal direction.

Charge exchange occurs provided the electron is able to cross this potential 
barrier. 
Let $N_{\Omega}$ be the 
fraction of trajectories which lead to electron loss at the time $t$. It is clear from
the discussion above that it must be  function 
of the solid opening angle angle $\Omega$, whose projection on the plane is the 
$\pm \theta_m$ angle. The exact expression for $N_{\Omega}$ will be given below.
Further, be $W(t)$ the probability for the electron to be still bound to the 
target, always at time $t$. Its rate of change is given by   
\begin{equation}
\label{eq:losses}
dW(t) = - N_{\Omega} dt {2 \over T_{em}} W(t) \quad , 
\end{equation}
with $T_{em}$ the period of the electron motion along its orbit.\\
It is important to discuss the factor $ dt (2/T_{em})$ since it is an important difference 
with (I), where just half of this value was used. The meaning of this factor is to account for
the fraction of electrons which, within the time interval $ [t, t + dt]$ reach and cross the potential
saddle. In (I) it was guessed that it should be equal to $dt/T_{em}$, on the basis of 
an uniform distribution of the classical phases of the electrons. However, let us read again 
what the rhs of eq. (\ref{eq:losses}) does mean: it says that the probability of loss
is given by the total number of {\it available} electrons within the loss cone
($ W(t) \times N_{\Omega}$), multiplied  by the fraction of electrons which reach the potential 
saddle. However, on the basis of the no--return hypothesis, only outgoing electrons can
contribute to this term: an electron which is within the loss cone and is returning to 
the target from the projectile is not allowed, it should already have been captured 
and therefore would not be in the set $W$. It is clear, therefore, that the effective
period is  $T_{em}/2$, corresponding to the outgoing part of the trajectory. 

A simple integration yields the leakage probability
\begin{equation}
\begin{split}
\label{eq:prob}
P_l =  P(+\infty) &=  1 - W(+\infty) = \\
      &=1 - \exp \left( - {2 \over T_{em}} \int_{- t_m}^{+ t_m} N_{\Omega} dt \right) \quad .
\end{split}
\end{equation}
In order to actually integrate Eq. (\ref{eq:prob}) we need to know the 
collision trajectory; an unperturbed straight line with $b$ impact parameter 
is assumed:
\begin{equation}
\label{eq:traiettoria}
R = \sqrt{b^{2} + (v t)^{2}} \quad .
\end{equation}
The extrema $ \pm t_m$ in the integral (\ref{eq:prob}) are the maximal values of $t$ at which
charge exchange can occur. If we identify this instant with the birth of the opening,
using eq. (\ref{eq:rm}) and (\ref{eq:traiettoria}), we find 
\begin{equation}
\label{eq:tc}
t_m = { \sqrt{R_m^2 - b^2} \over v} \quad .
\end{equation}

At this point it is necessary to give an explicit expression for 
$N_{\Omega}$. To this end, we will consider first the case of an electron with
zero angular momentum ($ l = 0$), and then will extend to nonzero values.

In absence of the projectile, the classical electron trajectories, with zero angular 
momentum, are ellipses squeezed onto the target nucleus. 
We are thus considering an electron moving essentially in one dimension. Its
hamiltonian can be written as
\begin{equation}
\label{eq:unod}
{p^2 \over 2} - {1 \over r} = - E_n \quad .
\end{equation}
The electron has a turning point at 
\begin{equation}
\label{eq:rc}
 r_c = { 1 \over E_n} \quad .
\end{equation}
Obviously the approaching of the projectile modifies these trajectories.
However, in order to make computations feasible, we make the following 
hypothesis: electron trajectories are considered as essentially unperturbed 
in the region between the target and the saddle point.
The only trajectories which are thus allowed to escape are those whose aphelia
are directed towards the opening within the solid angle whose projection
on the ${\cal P}$ plane is $\pm \theta_{m}$ (see fig. \ref{fig:opening}) {\it provided that} 
the turning point of the electron is greater than the saddle-point distance: $ r_c \geq z_0$. 
The validity of these approximations can be questionable, particularly 
if we are studying the collision with highly--charged ions, which could 
deeply affect the electron trajectory. We limit to observe that it is necessary 
in order to make analytical calculations. {\it A posteriori}, we 
shall check the amount of error introduced by such an approximation. 

The angular integration is now easily done, supposing a uniform distribution for the 
directions of the electrons:
\begin{equation}
N_\Omega = {1 \over 2} (1 - \cos \theta_m) \quad .
\end{equation}
In order to give an expression for $\theta_m$ we notice that
$ \cos\theta_m = z_{0} / (\rho_m^{2}  + z_0^{2})^{1/2}$, with $\rho_m$ root of
\begin{equation}
\label{eq:radici}
E(R)  = \left( \rho_m^2 + { R^2 \over (\sqrt{Z_p} + 1)^2 } 
\right)^{-1/2} + Z_p  \left( \rho_m^2 + { Z_p R^2 \over (\sqrt{Z_p} + 1)^2 }
\right)^{-1/2} \quad .
\end{equation}
It is easy to recognize that, in the right-hand side, the first term is 
the potential due to the electron--target interaction, and the second is 
the electron--projectile contribution.
Eq. (\ref{eq:radici}) cannot be solved analytically for $\rho_m$ except for the particular
case $Z_p = 1$, for which case:
\begin{equation}
\label{eq:rhouno}
\rho_m^2 = \left({ 2 \over E(R)}\right)^2 - \left({ R \over 2}\right)^{2}  \quad .
\end{equation}
The form of $E(R)$ function of $R$ cannot be given analytically, 
even though can be quite easily computed numerically \cite{cpc}. In 
order to deal with expressions amenable to algebraic manipulations, we 
do therefore the approximation: first of all, divide the space in the 
two regions $ R < R_{u}, R > R_{u}$, where $R_{u}$ is the 
internuclear distance at which the energy given by eq. (\ref{eq:due}) 
becomes comparable with its united--atom form:
\begin{equation}
\label{eq:rlimit}
E_n + {Z_p \over R_u} = (Z_p + 1)^2 E_n \rightarrow 
R_u =  { Z_p \over  (Z_p + 1)^2 - 1  }{ 1 \over E_{n} } \quad .
\end{equation} 
We use then for $E(R)$ the united--atom form for $R < R_{u}$, and the 
asymptotic form otherwise:
\begin{equation}
\begin{split}
\label{eq:ens}
E(R) &= E_{n} + {Z_{p} \over R} , \qquad R > R_{u} \\
     &= (Z_{p} + 1)^{2} E_{n} , \quad R < R_{u}
\end{split}
\end{equation}
It is worthwhile explicitly rewriting eq. (\ref{eq:rhouno}) for the 
two cases:
\begin{equation}
\begin{split}
\rho_{m}^{2} &=  R^{2} \left( {4 \over (E_{n} R + 1)^2 } - {1 \over 4}\right), 
                                \quad R >  R_{u} \\
         &= {1 \over 4} \left( { 1 \over E_{n}^2} - R^2 \right), \quad R < R_u 
\end{split}
\end{equation}
and the corresponding expressions for $N_{\Omega}$ are:
\begin{equation}
\begin{split} 
N_{\Omega} = {1 - \cos \theta_{m} \over 2 } &= {1 \over 8} ( 3 - E_{n} R) ,\quad  R > R_{u} \\
  &= {1 \over 2} ( 1 - E_{n} R) ,\quad R < R_{u} \quad .
\end{split}
\end{equation}
Note that $N_{\Omega} = 1/2 $ for $R = 0$. This is a check on the correctness of the model,
since, for symmetrical scattering at low velocity and small distances we expect the electrons 
to be equally shared between the two nuclei.\\

When $Z_{p} > 1$ we have to consider two distinct limits: 
when $ R \to \infty$ we know that eventually $\rho_{m} \to 0$ (eq. \ref{eq:rm}).
It is reasonable therefore to expand (\ref{eq:radici}) in series of powers of $\rho_m/R$
and, retaining only terms up to second order:
\begin{equation}
\label{eq:rhosol}
\rho_m^2 \approx { 2 \sqrt{Z_{p}} \over \left( \sqrt{Z_{p}} +  1\right)^{4} } R^{2} 
\left[ \left( \sqrt{Z_{p}} +  1\right)^{2}  - Z_{p} - E_{n} R \right]  \quad.
\end{equation}
Consistently with the limit $R \to \infty$, we have used the 
large--$R$ expression for $E(R)$. \\
The limit $ R \to 0$ is quite delicate to deal with: a straightforward 
solution of eq. (\ref{eq:radici}) would give 
\begin{equation}
\rho_m \approx {1 \over (Z_{p} + 1) E_{n}} + {\cal O}(R) \quad ,
\end{equation}
but calculating $\cos \theta_{m}$ and eventually $N_{\Omega}$ from 
this expression gives wrong results: it is easy to work out 
the result $N_{\Omega} = 1/2, R \to 0$.  This is wrong because, 
obviously, the limit $ N_{\Omega} \to 1, Z_{p} \to \infty$ must hold.
The reason of the failure lies in the coupling of eq. 
(\ref{eq:radici}) with  the united--atom form for $E(R)$: one can 
notice that the expression thus written is perfectly simmetrical 
with respect to the interchange projectile--target. Because of this 
symmetry, electrons are forced to be equally shared between the two nuclei.
This is good when dealing with symmetrical collisions, 
$ Z_{p} = Z_{t} = 1$, and is actually an improvement with respect to (I), 
where eq. (\ref{eq:rhosol}) was used even for small $R$'s and one recovered 
the erroneous value $N_{\Omega}(R = 0) = 3/8$.
But when $Z_{p} >1$ the asymmetry must be retained in the equations. The 
only way we have to do this is to extend eq. (\ref{eq:rhosol}) to 
small $R$, obtaining
\begin{equation}
\label{eq:costheta}
1 - \cos \theta_m \approx { \sqrt{Z_p} \over (\sqrt{Z_p} + 1)^2}
\left[  (\sqrt{Z_p} + 1)^2  - Z_p - E_n R \right] \quad .
\end{equation}
It is straightforward to evaluate eq. (\ref{eq:costheta}) 
in the limit $Z_{p} \to \infty, R \to 0$, and find the sought result, 2.\\
We notice that, from the numerical point of view, it is not a great error
using eq. (\ref{eq:rhosol}) everywhere: 
the approximation it is based upon breaks down when $R$ is of 
the order of $R_{u}$ or lesser, which is quite a small range with 
respect to all other lengths involved when $Z_{p} > 1$, while even for 
the case $Z_p = 1$ it is easy to recover (see equations below) 
that the relative error thus introduced 
on $P_l$ is $ \Delta P_l/P_l = 1/24$ for small $b$ (and--obviously--it is
exactly null for large $b$). Therefore, eq. (\ref{eq:rhosol}) could be 
used safely in all situations.
However, we think that the rigorous altough quite lengthy derivation given 
above was needed since it is not satisfactory working with a model 
which does not comply with the very basic requirements required by the symmetries 
of the problem at hand. \\
 
We have now to take into account that the maximum escursion for the electron
is finite. If we put $r_c = z_0$ and use for $z_{0}$, $r_c$ respectively
the expressions given by (\ref{eq:saddle}) and (\ref{eq:rc}), we obtain an
equation which can be easily solved for $R$:
\begin{equation}
\label{eq:rmp}
R = {R'}_m = { (\sqrt{Z_{p}} + 1 )  r_c} \quad .
\end{equation}
The ${R'}_m$ thus computed is the maximum internuclear distance at which charge
exchange is allowed under the present assumptions.
Since ${R'}_m < R_m$ (compare the previous result with that of eq. \ref{eq:rm} )
we have to reduce accordingly the limits in
the integration in eq. (\ref{eq:prob}): it must be performed between 
$ \pm {t'}_m$, with the definition of ${t'}_m$ the same as $t_m$ but for the replacement 
$ R_m \to {R'}_m$. \\
The result for the leakage probability is:
\begin{equation}
\label{eq:probexplicit}
P_l = 1 - \exp \left( - 2 {F(u_m) + G_{Z} \over T_{em}} \right) \quad ,
\end{equation}
where we have defined
\begin{equation}
\begin{split}
\label{eq:effe}
F(u)  &= { \sqrt{Z_p}   \over  (\sqrt{Z_p} + 1)^2 }
\left[  \left( (\sqrt{Z_{p}} + 1)^{2} - Z_{p} \right) { b \over v} u -  
\left({E_n b^{2} \over 2 v}\right) \left( u \sqrt{ 1 + u^{2}} + 
{\rm arcsinh}(u) \right) \right] \quad ,\\
G_Z &= (3 F(u_u) - 2 t_u) \quad (Z_{p} = 1) \\
    &= 0 \quad (Z_{p} > 1) \quad , \\
u_m &= v t'_{m}/b  \quad ,\\
u_u &= v t_u/b  \quad ,\\
t_u &= { \sqrt{R_u^2 - b^2} \over v}  \quad .
\end{split}
\end{equation}
The period can be easily computed by
\begin{equation}
T_{em} = 2 \int_0^{1/E_n}{dr \over p} = 
 \sqrt{2} \int_0^{1/E_n} {dr \over \sqrt{ {1 \over r} - E_n}} = 2 \pi n^3 
\end{equation}
(this result could be found also in \cite{landau}).\\
The cross section can be finally obtained after integrating over the impact
parameter (this last integration must be done numerically):
\begin{equation}
\label{eq:sigma}
\sigma = 2 \pi \int^{b_m}_0 b P_l(b) db \quad .
\end{equation}
Again, we have used the fact that the range of interaction is finite: the maximum
allowable impact parameter $ b_m$ is set equal to ${R'}_m$. \\

Finally, we consider the case when the angular momentum is different from 
zero. Now, orbits are ellipses whose minor
semiaxis has finite length. We can still write the hamiltonian as function of just $(r,p)$: 
\begin{equation}
\label{eq:lnotzero}
{p^2 \over 2} - {1 \over r}  + {L^2 \over 2 r^2}= - E_n \quad .
\end{equation}
$L$ is the usual term: $ L^{2} = l (l + 1)$.
The turning points are now 
\begin{equation}
\label{eq:rc2}
 r_c^{\pm} = { 1 \pm \sqrt{1 - 2 E_n L^2} \over 2 E_n} \quad .
\end{equation}
and $R'_{m} = (\sqrt{Z_p} + 1) r^{+}_{c}$.

Now the fraction of trajectories entering the loss
cone is much more difficult to estimate. In principle, it can still be determined: 
it is equal to the fraction of ellipses which have intersection with the opening. 
Actual computations can be rather cumbersome. Thus, we use the following approximation, 
which holds for low angular momenta $ l << n$ (with $n$ principal quantum number): 
ellipses are approximated as straight lines (as for the $l = 0$ case), but their turning 
point is correctly estimated using eq. (\ref{eq:rc2}).
Note that also the period is modified: its correct expression is
\begin{equation}
T_{em} =  
  \sqrt{2} \int_{r^-}^{r^+} {dr \over \sqrt{ {1 \over r} - E_n - {l (l+1) \over 2 r^2}}} 
\quad .
\end{equation}
  
\section{A test case}
As a first test case we consider the inelastic scattering
${\rm Na}^+ + {\rm Na(28d, 29s)}$.
We investigate this sytem since: (i) it has been studied experimentally in \cite{nana};
(ii) some numerical simulations using the Classical Trajectory Monte Carlo
(CTMC) method have also been done on it \cite{pascale}, allowing to have
detailed informations about the capture probability $P_l$ function
of the impact parameter, and not simply integrated cross sections;
(iii) finally, it has been used as test case in (I), thus allowing
to assess the relative quality of the fits.

In fig. (\ref{fig:nana1}) we plot the normalized cross section $\tilde{\sigma} = \sigma/n^4$ 
{\it versus } the normalized impact velocity $\tilde{v} = v n$ for both
collisions $nl = $ 28d and $nl = $ 29s (solid line).
The two curves are very close to each other, reflecting the fact that the two orbits
have very similar properties: the energies of the two states differ by a 
very small amount, and in both cases $ E_n L^2 << 1$. The two
curves show reversed with respect to experiment: $\sigma$(28d) it is greater
than $\sigma$(29s). The reason is that the parameter $r_c$ is larger in the former 
case than in the latter.\\ 
We can distinguish three regions: the first is at reduced velocity around 0.2, where a steep
increase of cross section appears while going towards lower velocities. Over--barrier models
do not appear to fully account for this trend: they have a behaviour at low speed which
is ruled approximately by the $1/v$ law, consequence of the straight-line impact trajectory 
approximation: it is well possible that this approximation too becomes unadequate in 
this region. \\
The second region covers roughly the 
range 0.3 $\div$ 1.0. Here the $nl = $ 29s data are rather well simulated while the 
present model overestimates the data for $ nl = $ 28d. The bad agreement for
$nl = $ 28d was already clear to Ostrovsky which attributed it to a deficiency of the 
model to modelize $l$-changing processes. It seems clear that neither our treatment
of the angular momentum is sufficient to cure this defect.

Finally, there is the region at $\tilde{v} > 1$, where again the OBM, as it stands, 
is not able to correctly reproduce the data. The reason for this discrepancy 
can be traced back to the finite velocity of the electron: 
the classical electron velocity is $ v_e = 1/n$, so $\tilde{v}$ can be given 
the meaning of the ratio between the projectile and the electron velocity. 
When $\tilde{v} \geq 1$ the projectile
is less effective at collecting electrons in its outgoing part of the trajectory 
(i.e. when it has gone beyond the point of closest approach). In simple 
terms: an electron is slower than the projectile; when it is left behind, it cannot 
any longer reach and cross the potential barrier.

\section{Corrections to the model}
\begin{figure}
\epsfxsize=12cm
\epsfbox{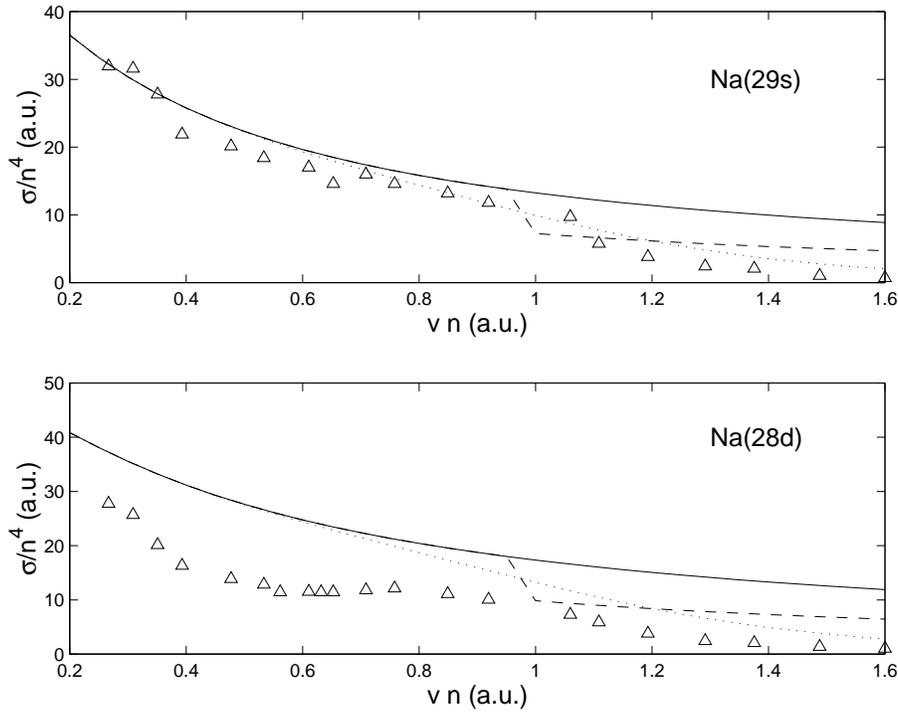}
\caption{Cross section for charge exchange for  Na${}^{+}$--Na(29s)
(upper) and  Na${}^{+}$--Na(28d) (lower) collisions. Symbols, experimental data 
(adapted from ref. 7); solid line, present model with $w^{+} = 1$;
dashed line, model with $w^{+} = 0$; dotted line, model with $w$ 
given by eq. (32).  
Note that the experimental results are not absolutely calibrated, the data shown 
here are calibrated using as reference the CTMC results at $ \tilde{v} = 1$ 
and $ nl = $ 28d.}
\label{fig:nana1}
\end{figure}

This picture suggests a straightforward remedy: a term must be inserted in eq. (\ref{eq:losses})
to account for the diminished capture efficiency. This is accomplished formally through 
rewriting $ N_\Omega \to w(t,\tilde{v}) N_\Omega$, with $ w \leq 1$. We have put into
evidence that $w$ can in principle be function of time and of the impact velocity.
The simplest correction is made by assuming a perfect efficiency for $\tilde{v} < 1$,
$ w(t, \tilde{v} < 1) = 1$, while, for $\tilde{v} > 1$, no electrons can be collected after 
that the distance of minimum approach has been reached: 
$ w^+ \equiv w(t > 0, \tilde{v} > 1) = 0$. This can appear too strong an 
assumption, since those electrons which are by the same side of the projectile 
with respect to the nucleus, and which are close to their turning point may still be captured. 
In fig. (\ref{fig:nana1}) we can compare the original data with those   
for $ w^+ = 0$ (dashed line). The sharp variation of $\sigma$ at $\tilde{v} = 1$
is obviously a consequence of the crude approximations done choosing $w$ which 
has a step--like behaviour with $v$. 

\begin{figure}
\epsfxsize=12cm
\epsfbox{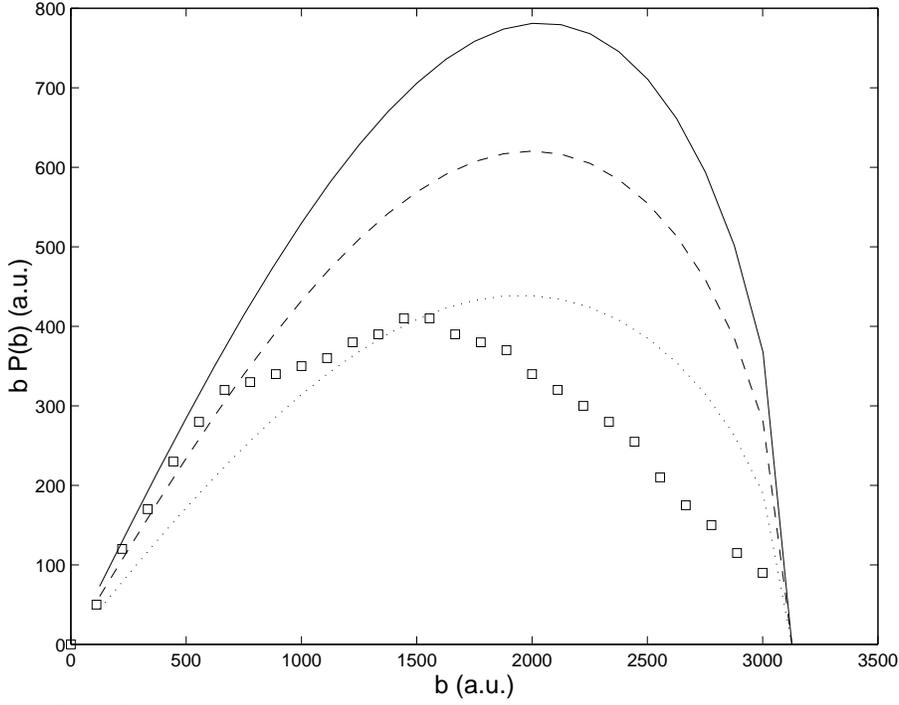}
\caption{Probability of electron capture multiplied by impact 
parameter, $P_{l } b$, for Na${}^{+}$--Na(28d) collision at 
$\tilde{v} = 1 $. Squares, CTMC data (adapted from ref. 8); 
solid line, present model with $w^{+} = 1$; dashed line, 
$w^{+} = 0.5$; dotted line, $w^{+} = 0$.}
\label{fig:nana}
\end{figure}

To get further insight, we plot in fig. \ref{fig:nana} the quantity 
$b P_l(b)$ {\it versus} $b$ for the collision ${\rm Na}^+ + {\rm Na(28d)}$. 
The impact velocity is $ \tilde{v} = 1 $.
The symbols are the CTMC results of ref. \cite{pascale}. Solid line is 
the model result for $w^{+} = 1$; dotted line, the result for $w^{+} = 
0$; dashed line, an intermediate situation, with $w^{+} = 1/2$.
Striking features are, for all curves, the nearly perfect accordance of 
the value $b \approx 3000$ at which $P_l = 0$ (it is $b_m$
according to our definition). The behaviour at small $b$'s ($ P_l \approx 1/2$)
is well reproduced for $w^{+} = 1$ while it is slightly underestimated 
by the two other curves. On the other hands, only by setting $w^{+} = 
0$ it is possible to avoid the gross overestimate of 
$P_l$ near its maximum. 

It is thus evident that the agreement is somewhat improved in the region  
$\tilde{v} \approx 1$ by letting $w^{+} = 0$.
However, the high--velocity  behaviour is still missed by the model, which predicts a 
power--law behaviour $\sigma \propto v^{-1}$, while the actual exponent is higher.   
Within our picture, this suggests that also the capture efficiency 
$w^- = w(t < 0)$ must be a decreasing function of $\tilde{v}$. 
An accurate modelization of the processes which affect this term is difficult,
and we were not able to provide it. However, some semi--qualitative arguments can
be given. Let us review again the process of capture  
as described in section II and shown in fig. (\ref{fig:opening}):
if  $\tilde{v} > 1$, an electron at time $t$ can be in the loss cone and still not to 
be lost, since
within a time span $\Delta t \approx \rho_{m}/v$ the position of the loss cone has 
shifted of such an amount that  only those electrons which were closer to the saddle point 
than a distance $ v_{e} \Delta t$ could be caught. The fraction of these 
electrons is 
$ \Delta t (2 /T_{em}) \approx  \rho_{m} ( 2 /v T_{em}) $.  This correction gives an additional
$1/v$ dependence, thus now $\sigma \approx 1/v^2$.\\
As an exercise, we try to fit experimental data using $w$ as a free parameter instead that
a function to be determined by first principles.
We choose one of the simplest functional forms:
\begin{equation}
\label{eq:formaw}
 w = {1 + |\beta|^m \over  1 + | \tilde{v} - \beta|^m} \quad ,
\end{equation}
with $\beta, m$ free parameters to be adjusted. This form gives the two correct limits:
$ w \to 1, \tilde{v} \to 0$, and $ w \to 0, \tilde{v} \to \infty$. The parameter
$\beta$ is not really needed; it has been added
to reach a better fit. Its meaning is that of a treshold velocity, at which
the capture efficiency begins to diminish. In fig. (\ref{fig:nana1}) we plot 
the fit obtained with $\beta = 0.2, m = 4$ (dotted line): this is not meant to be the best fit, 
just a choice of parameters which gives a very good agreement with data.  We see that 
the suggested corrections are still not enough to give the right 
power--law, if one needs to go to some extent beyond the region $ \tilde{v} = 1$.

\section{Other comparisons}

\subsection{Iodine - Cesium collisions}
\begin{figure}
\epsfxsize=12cm
\epsfbox{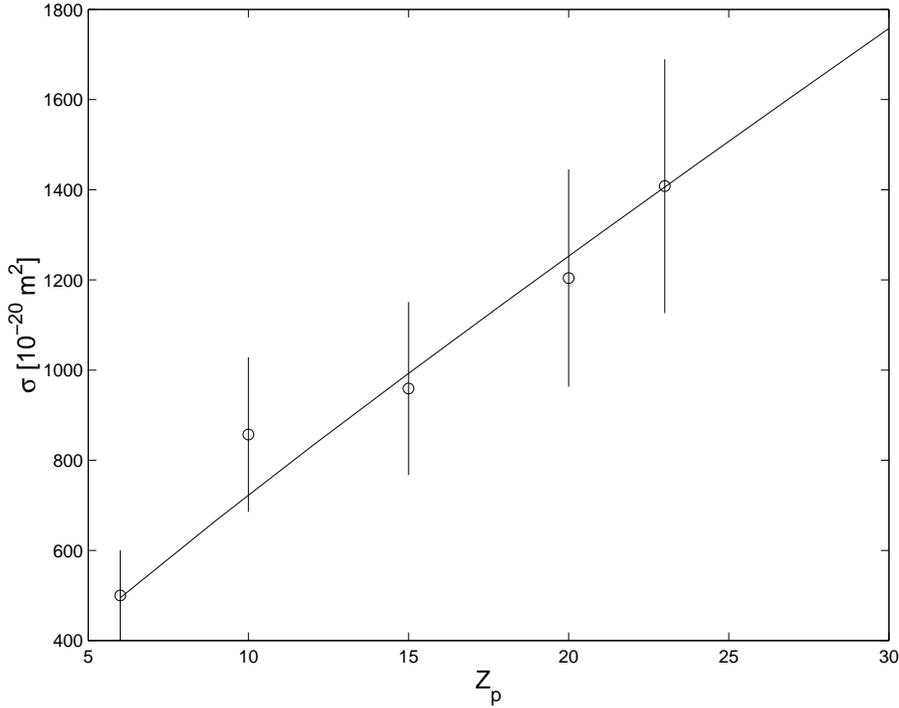}
\caption{Cross section for charge exchange in I${}^{+q}$--Cs collisions.
Circles, experimental data with 20\% error bar; solid line, present 
model (where we have set $ w \equiv 1$, since we are dealing with 
$ v/v_e << 1$).}
\label{fig:ics}
\end{figure}

We apply now our model to the process of electron capture
\begin{equation}
\label{eq:processo}
{\rm I}^{q+} + {\rm Cs} \to {\rm I}^{(q-1)+} + {\rm Cs}^{+}
\end{equation}
with $q = 6 \div 30$. This scattering process has been studied experimentally
in \cite{sette}. It is particularly interesting to study in this context 
since it has revealed untractable by a number of other OBM's, including that 
of (I) (for a discussion and results, see \cite{jpb}). 
The impact energy is chosen equal to $1.5\times Z_p$ keV: since it corresponds
to $\tilde{v} <<1$, we can safely assume $w = 1$.
The Cesium atom is in its ground state with the optical electron in a 
$s$ state. \\
In fig. \ref{fig:ics} we plot the experimental points together with our estimates. In this 
case the fit is excellent.
It is important to notice that this agreement is entirely consequence of
our choice of limiting integration to $R$ given by eq. (\ref{eq:rmp}): to understand
this point, observe that because of the very high charge of the projectile, the 
exponential term in eq. (\ref{eq:probexplicit}) is small ($F$, by direct inspection,
is increasing with $Z_p$) and thus $P_l \approx 1 $. The details of the model
which are in $F$ are therefore of no relevance. The only surviving parameter, and that which
determines $\sigma$, is $R'_m$. It can be checked by directly comparing 
our fig. \ref{fig:ics}
with fig. 1 of ref. \cite{jpb}, where results from model (I) are shown, 
which differ from ours just in replacing eq. 
(\ref{eq:rmp}) with eq. (\ref{eq:rm}). There, the 
disagreement is severe.
   
\subsection{Ion - Na($n = 3 $) collisions}
\begin{figure}
\epsfxsize=12cm
\epsfbox{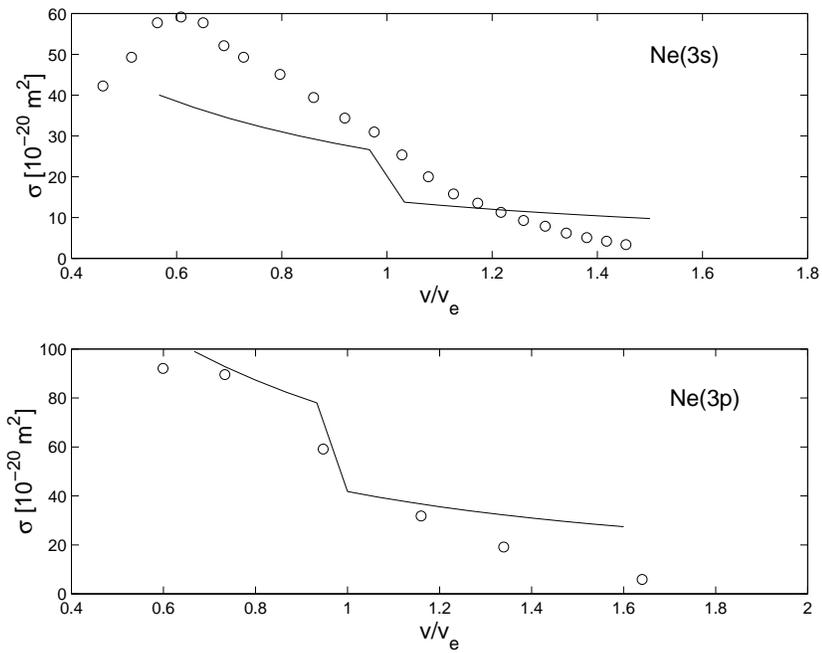}
\caption{Cross section for charge exchange in H${}^{+}$--Na(3s)
(upper) and  H${}^{+}$--Na(3p) (lower) collisions. 
Symbols, experimental data from ref. (9); lines, present model.}
\label{fig:ionna}
\end{figure}

As a final test case we present the results for collisions H--Na(3s,3p).
They are part of a set of experiments as well as numerical simulations
involving also other singly--charged ions: He, Ne, and Ar (see 
\cite{na3} and the references therein and in particular \cite{na2};
ref. \cite{na4} presents numerical calculations for the same system). 
In fig. \ref{fig:ionna} we plot the results of our model together with
those of ref. \cite{na3}. Again, we find that only by neglecting $w^+$ 
some accordance is found. 
The low--energy wing of the curve is strongly underestimated for Na(3s), while 
the agreement is somewhat better for Na(3p). Again, the slope of $\sigma$ for
relative velocities higher than 1 could not be reproduced. \\ 
We do not show results for other ions: they can be found in fig. 3 of ref. \cite{na3}.
What is important to note is that differencies of a factor two (and even larger for 3s states)
appear between light (H${}^+$, He${}^+$) and heavy (Ne${}^+$, Ar${}^+$) ions
which our model is unable to predict.
We can reasonably conclude therefore: (i) that the present model is not 
satisfactory for $ v/v_e << 1$ (it was already pointed out in sec. IV) and for 
$ v/v_e > 1$ ;
(ii) the structure of the projectile must be incorporated into the 
model otherwise different ions with the same charge should cause the same effect,
at odds with experiments. As emphasized in \cite{na3,na2} the 
energy defect $\Delta E$ of the process is a crucial parameter: captures to states 
with $\Delta E \approx 0$ are strongly preferred. Obviously, the value 
of $\Delta E$ depends on the energy levels structure of the 
recombining ion.

\section{Summary and conclusions}
We have developed in this paper a classical OBM for single charge 
exchange between ions and atoms. 
The accuracy of the model has been tested against three cases, with
results going from moderate--to--good (sec. III and IV), excellent (sec. V.A), 
and poor--to--moderate (sec. V.B). As a rule of thumb, the model can be stated 
to be very well suited for collisions involving highly charged ions at low 
velocities.\\
The model is based upon a previous work \cite{ostrovsky}, and adds 
to it a number of features, which we go to recall and discuss: 
(i) the finite excursion from the nucleus permitted to the electrons; 
(ii) the redefinition of the fraction of lost electrons
$ dt/T_{em} \to  dt (2/T_{em} )$; 
(iii) a more accurate treatment of the small impact parameter region for
symmetrical collisions; 
(iv) the explicit-altough still somewhat approximate-treatment of the capture 
from $ l > 0$ states;
(v) a correction to the capture probability due finite impact velocity. Let us discuss 
briefly each of these points:\\
Point (i) and (ii) contribute a major correction: in particular, (i) is essential to
recover that excellent agreement found in section V.A,  while (ii) accounts for the 
correct $b P_l$ behaviour at small $b$'s (see fig. 2).\\
Point (iii) is unimportant for actual computations, but corrects an
inconsistency of the model. \\   
Point (iv) has been studied in less detail, in part for the lack of experimental 
data on which doing comparisons. \\
Point (v): a good theoretical estimate of $w$  
should be of the outmost importance for developing a really accurate 
model of collision at medium-to-high impact velocity. 
In this paper we have just attempted a step towards this direction 
which, however, has allowed to recover definitely better results. \\
Finally we recall from sec. V.B that the treatment of the 
projectile--or better the process of the electron-projectile 
binding--is an aspect which probably awaits for main improvements. 
We just observe that it is a shortcoming of all classical methods,
that they cannot easily deal with quantized energy levels.

\section*{Acknowledgments}
It is a pleasure to thank the staff at National Institute for Fusion 
Science (Nagoya), and in particular Prof. H. Tawara and Dr. K. Hosaka 
for providing the data of ref. \cite{sette}.


\newpage


\end{document}